\documentclass[letterpaper,11pt]{article}
\pdfoutput=1 
\usepackage{fullpage}
\usepackage{amsmath}
\usepackage{amssymb}
\usepackage{graphicx}
\usepackage{mathrsfs}
\usepackage{wrapfig}
\usepackage{boxedminipage}
\usepackage{epsfig}
\usepackage{setspace}
\usepackage{filecontents}
\usepackage{subcaption}
\usepackage{float}
\usepackage{hyperref}
\usepackage{enumerate}
\usepackage{cite}
\usepackage[font=footnotesize,labelfont=rm]{caption}

\newcommand{\beq}{\begin{equation}}
\newcommand{\eeq}{\end{equation}}
\newcommand{\bea}{\begin{eqnarray}}
\newcommand{\eea}{\end{eqnarray}}

\begin{document}
\begin{flushright}
{\tt arXiv:1903.00613
}
\end{flushright}

\bigskip
\bigskip
\bigskip
\bigskip

\bigskip
\bigskip
\bigskip
\bigskip

\begin{center}
{\Large
{\bf On the Complexity of a $2+1$--dimensional Holographic Superconductor
}
}
\end{center}

\bigskip
\begin{center}
{\bf  Avik Chakraborty }

\end{center}

\bigskip
\bigskip

\begin{center}
  \centerline{\it Department of Physics and Astronomy }
\centerline{\it University of
Southern California}
\centerline{\it Los Angeles, CA 90089-0484, U.S.A.}
\end{center}

\bigskip
\centerline{\small \tt avikchak,  [at] usc.edu}

\bigskip
\bigskip
\bigskip

\begin{abstract} 
\noindent  
We present the results of our computation of the subregion complexity and also compare it with the entanglement entropy of a $2+1$--dimensional holographic superconductor which has a fully backreacted gravity dual with a stable ground sate. We follow the ``complexity equals volume'' or the CV conjecture. We find that there is only a single divergence for a strip entangling surface and the complexity grows linearly with the large strip width. During the normal phase the complexity increases with decreasing temperature, but during the superconducting phase it behaves differently depending on the order of phase transition. We also show that the universal term is finite and the phase transition occurs at the same critical temperature as obtained previously from the free energy computation of the system. In one case, we observe multivaluedness in the complexity in the form of an ``S'' curve.
\end{abstract}
\newpage \baselineskip=18pt \setcounter{footnote}{0}

\section{Introduction}
\label{Intro} 
 
The AdS/CFT correspondence provides us a dual description of the $d$--dimensional strongly interacting field theories on the boundary and the $d+1$--dimensional weakly coupled gravity theories in the bulk \cite{Maldacena:1997re, Aharony:1999t}. This correspondence has been used extensively in many contexts over the past decade. Two quantities on the boundary field theory play important roles in the quantum information theory: the entanglement entropy and the computational complexity. Surprisingly, both these quantities are reflected in the bulk geometry.

The entanglement entropy is a measure of the quantum correlations of a quantum state and is extremely useful for studying black hole physics, condensed matter systems etc. Given a quantum mechanical system, let us divide it into two subsystems: $\mathcal{A}$ and its complement $\mathcal{B}=\bar{\mathcal{A}}$. Then the reduced density matrix of $\mathcal{A}$ is given by taking the partial trace over the system $\mathcal{B}$: $\rho_{\mathcal{A}}=\mathrm{Tr}_{\mathcal{B}}(\rho_{tot})$, where $\rho_{tot}$ is the total density matrix of the system. Then the entanglement entropy of the subsystem $\mathcal{A}$ is defined as the von Neumann entropy of the reduced density matrix $\rho_{\mathcal{A}}$: $S_{\mathcal{A}}=-\mathrm{Tr}_{\mathcal{A}}(\rho_{\mathcal{A}}\log \rho_{\mathcal{A}})$. The Ryu-Takayanagi (RT) conjecture tells us how to compute the entanglement entropy holographically \cite{Ryu:2006bv, Ryu:2006ef}. To be more specific, it tells that the holographic entanglement entropy of a subregion $\mathcal{A}$ with its complement in the $d$--dimensional boundary is given by:
\bea \label{eeform}
\mathcal{S}=\frac{{\rm Area}(\gamma_{\mathcal{A}})}{4G_{d+1}} \ ,
\eea
where $\gamma_{\mathcal{A}}$ is the extremal surface in the bulk, that extends from the boundary of the region $\mathcal{A}$ and $G_{d+1}$ is the Newton's constant in the $d+1$--dimensional bulk.

On the other hand, the computational complexity of a quantum state can be roughly interpreted as the minimum number of gates required to implement a certain unitary operator to prepare this state from a given reference state. A precise definition of the complexity in the boundary CFT remains an open problem and recently there have been many attempts to address this question. Studies of circuit complexity of Gaussian states in the free field theories using Nielsen's geometric approach and Fubini-Study metric have been investigated in refs. \cite{Chapman:2017rqy, Jefferson:2017sdb, Khan:2018rzm, Chapman:2018hou, Ali:2018fcz, Ali:2018aon}. A path-integral optimization procedure to define the computational complexity is explored in refs. \cite{Caputa:2017urj, Caputa:2017yrh, Bhattacharyya:2018wym, Takayanagi:2018pml}. Recently, there have been two conjectures to compute the complexity in holography \cite{Susskind:2014moa, Susskind:2014rva, Brown:2015bva, Brown:2015lvg}. The first conjecture is known as the ``complexity equals volume'' or CV. It says that for an eternal black hole the complexity is proportional to the spatial volume of the Einstein-Rosen bridge connecting two boundaries. The second one is known as the ``complexity equals action'' or CA which relates the complexity to the action on a Wheeler-DeWitt patch\footnote{For a selection of references see \cite{Carmi:2016wjl, Couch:2016exn, Huang:2016fks, Cai:2016xho, Cai:2017sjv, Fu:2018kcp, Alishahiha:2017hwg, Momeni:2017kbs, Alishahiha:2018tep, HosseiniMansoori:2018gdu, Alishahiha:2018lfv, Ghodrati:2017roz, Ghodrati:2018hss} where the authors have developed and further extended these ideas.}. In this paper we will focus only on the CV conjecture. Based on the CV conjecture, Alishahiha in \cite{Alishahiha:2015rta} has proposed that the holographic complexity of a subregion $\mathcal{A}$ is equal to the codimension-one maximal volume of the bulk enclosed by the entangling region and the RT surface that appears in the holographic entanglement entropy computation:
\bea \label{cvform}
\mathcal{C}=\frac{{\rm Volume}(\gamma_{\mathcal{A}})}{8\pi R G_{d+1}} \ ,
\eea
where $R$ is the AdS radius. This quantity is known as the holographic subregion complexity. In \cite{Agon:2018zso, Caceres:2018blh} the subregion complexity has been explained as the purification complexity using tensor network model.

In the AdS/CFT framework, one important object which is widely investigated is the holographic superconductor \cite{Hartnoll:2008vx, Hartnoll:2008kx, Gubser:2009qm, Horowitz:2010gk, Aprile:2011uq, Salvio:2012at}. In recent years there has been numerous work carried out to model such a holographic superconductor dual to gravity theories coupled to a Maxwell field plus a scalar. In general the task is highly non-trivial and finding a stable vacuum is difficult. Recently, ref. \cite{Bobev:2011rv} has developed one such model in the context of AdS$_{4}$/CFT$_{3}$. Their solution is fully back reacted and the ground state is highly stable. Ref. \cite{Albash:2012pd} has further discussed about the phase transition of this system by studying the entanglement entropy using RT prescription. The action in ref. \cite{Bobev:2011rv} arises as an $SO(3)\times SO(3)$ invariant truncation of four--dimensional $\mathcal{N}=8$ gauged supergravity. One needs to numerically solve the equations of motion coming from this action. At high temperature, the solution is simply the RN--AdS solution. The scalar has zero value and hence there is no condensate. At low enough temperature, below some critical value a new type of solution exists with a non--zero charged scalar hair. This solution is thermodynamically preferred over the RN--AdS solution. Depending on the boundary conditions, there are two types of solutions: one gives rise to a second order phase transition and the other one first order phase transition. At zero temperature, the solution is an RG flow between two AdS spaces. Because the complexity measures the difficulty of turning one quantum state into another, it is expected that the subregion complexity should capture behavior of the phase transition of such a model and can provide useful information as well. In refs. \cite{Momeni:2016ekm, Zangeneh:2017tub, Fujita:2018xkl, Yang:2019gce, Guo:2019vni} the authors have discussed the subregion complexity for different types of holographic superconductors which have backreaction included and we will compare our results with theirs in section \ref{eehc} and \ref{sum}.

In this paper we will present our results of the subregion complexity of the $2+1$--dimensional holographic superconductor system mentioned above. We will see that there is a single divergence and no other divergences during the phase transition, the complexity remains finite during both the first order and the second order phase transitions. The complexity curves go linearly with $l$ for large strip width $l$. Moreover, the temperature where normal phase turns into a superconducting phase is exactly same in both the entanglement entropy and the subregion complexity computation and is equal to the transition temperature $T_c$ obtained from the free energy calculation in \cite{Albash:2012pd}. Another interesting fact that we will describe in section \ref{eehc} is that of the multivaluedness captured in different ways for both the cases. We will also see the discontinuous but finite jump behavior of the complexity for the first order phase transition. 

The outline of this paper is as follows. In the next section we will review the dual gravity background of our $2+1$--dimensional superconductor system, along with the solutions in different temperature regime. In section \ref{eehc} we will present our results for both the entanglement entropy and the subregion complexity of this system. We will conclude this paper with a short summary and discussions. In an appendix \ref{gen} we will briefly present some analytic results of the entropy and the complexity for general $d+1$--dimensional AdS black holes and compare them with our results in section \ref{eehc}.

\section{Dual Gravity Background of $2+1$--dimensional Holographic Superconductor}
\label{background}

In this section we will briefly review the dual gravity background of our $2+1$--dimensional holographic superconductor system. The Lagrangian that gives rise to this superconductor is \cite{Bobev:2011rv, Albash:2012pd}:
\bea \label{lag}
e^{-1}\mathcal{L}=\frac{1}{16 \pi G_{4}}\Big(\mathcal{R} - \frac{1}{4}F_{\mu \nu}F^{\mu \nu} - 2\partial_{\mu}\lambda \partial^{\mu}\lambda - \frac{\sinh^{2}(2\lambda)}{2}\big(\partial_{\mu}\varphi - \frac{g}{2}A_{\mu}\big)\big(\partial^{\mu}\varphi - \frac{g}{2}A^{\mu}\big) - \mathcal{P}\Big) \ \ ,
\eea
where the potential $\mathcal{P}$ is given by:
\bea
\mathcal{P}=-g^{2}\Big(6\cosh^{4}\lambda - 8\cosh^{2}\lambda \ \sinh^{2}\lambda + \frac{3}{2}\sinh^{4}\lambda \Big) \ .
\eea
This is the geometry of a black hole coupled with scalar fields $\lambda$ and $\varphi$ with a non-trivial potential, $A_{\mu}$ being the gauge field. $G_{4}$ is the $3+1$-dimensional gravitational constant. We choose the ansatz for the metric, the gauge field and the scalar field as follows:
\bea \label{ans}
ds^{2}=-\frac{R^{2}}{z^{2}}f(z)e^{- \chi (z)}dt^{2}+\frac{R^{2}}{z^{2}}(dx^{2}_{1}+dx^{2}_{2})+\frac{R^{2}}{z^{2}}\frac{dz^{2}}{f(z)} \ \ , \ \ A_{t}=\Psi (z) \ \ , \ \ \lambda = \lambda (z) \ ,
\eea
where we have set the scalar $\varphi$ to zero using the equations of motion and symmetry. Next, we define a dimensionless coordinate: $z=R\tilde{z}$. Then substituting this ansatz into the equations of motion arising from the Lagrangian (\ref{lag}) we get the following system of ordinary differential equations:
\begin{eqnarray}
&&\hskip -11.5cm \chi^{\prime}-2\tilde{z}(\lambda^{\prime})^{2}-\frac{\tilde{z}e^{\chi}\sinh^{2}(2\lambda)\Psi^{2}}{8f^{2}}=0 \label{chi} \ \ , \\
(\lambda^{\prime})^{2}-\Bigg(\frac{f^{\prime}}{\tilde{z}f}\Bigg)+\frac{\tilde{z}^{2}e^{\chi}(\Psi^{\prime})^{2}}{4f}+\frac{R^{2}\mathcal{P}}{2\tilde{z}^{2}f}+\frac{3}{\tilde{z}^{2}}+\frac{e^{\chi}\sinh^{2}(2\lambda)\Psi^{2}}{16f^{2}}=0 \label{eff} \ \ , \\
&&\hskip -11.5cm \Psi^{\prime \prime}+\Bigg(\frac{\chi^{\prime}}{2}\Bigg)\Psi^{\prime}-\frac{\sinh^{2}(2\lambda)\Psi}{4\tilde{z}^{2}f}=0 \label{psi} \ \ , \\
&&\hskip -11.5cm \lambda^{\prime \prime}+\Bigg(-\frac{\chi^{\prime}}{2}+\frac{f^{\prime}}{f}-\frac{2}{\tilde{z}}\Bigg)\lambda^{\prime}-\frac{R^{2}}{4\tilde{z}^{2}f}\frac{d\mathcal{P}}{d\lambda}+\frac{e^{\chi}\sinh(4\lambda)\Psi^{2}}{16f^{2}}=0 \label{lambda} \ \ .
\end{eqnarray}
The horizon is the zero locus of $f(\tilde{z})$. We will assume that it occurs at $\tilde{z}=\tilde{z}_{H}$. There are three scaling symmetries of the equations of motion\cite{Albash:2012pd}:
\begin{eqnarray}
&&\hskip -10.8cm t\rightarrow \gamma_{1}^{-1}t \ , \ \chi \rightarrow \chi -2\ln \gamma_{1} \ , \ \Psi \rightarrow \gamma_{1}\Psi \label{sym1} \ , \nonumber \\
&&\hskip -10.5cm t\rightarrow \gamma_{2}^{-1}t \ , \ z\rightarrow \gamma_{2}^{-1}z \ , \ R\rightarrow \gamma_{2}^{-1}R \label{sym2} \ , \nonumber \\
x^{\mu}\rightarrow \gamma^{-1}x^{\mu} \ , \ f(z)\rightarrow f(z) \ , \ \Psi(z)\rightarrow \gamma \Psi(z) \ , \ \lambda(z)\rightarrow \lambda(z) \ , \ \chi(z)\rightarrow \chi(z) \label{sym} \ .
\end{eqnarray}
Using these scaling symmetries we can choose arbitrary values of the position of the event horizon, the coupling constant of gauged supergravity, $g$, and the asymptotic value of the field $\chi(z)$. We choose the following:
\bea \label{scaling}
\tilde{z}_{H}=1 \ \ , \ \ g=1 \ \ , \ \ \lim_{z\rightarrow 0}\chi=0 \ \ .
\eea
To solve the equations of motion we need to know the IR and the UV behavior of various fields. Near the IR, i.e. the horizon, the fields have an expansion:
\begin{eqnarray}
&&\hskip -8.7cm \lambda(\tilde{z})=\lambda^{(0)}+\lambda^{(1)}\Bigg(1-\frac{\tilde{z}}{\tilde{z}_{H}}\Bigg)+... \ , \nonumber \\
&&\hskip -8.7cm \chi(\tilde{z})=\chi^{(0)}+\chi^{(1)}\Bigg(1-\frac{\tilde{z}}{\tilde{z}_{H}}\Bigg)+... \ , \nonumber \\
&&\hskip -8.7cm f(\tilde{z})=f^{(1)}\Bigg(1-\frac{\tilde{z}}{\tilde{z}_{H}}\Bigg)+... \ , \nonumber \\
\Psi(\tilde{z})=\Psi^{(1)}\Bigg(1-\frac{\tilde{z}}{\tilde{z}_{H}}\Bigg)+\Psi^{(2)}\Bigg(1-\frac{\tilde{z}}{\tilde{z}_{H}}\Bigg)^{2}+... \label{ir} \ .
\end{eqnarray}
Plugging this into the equations of motion (\ref{chi}-\ref{lambda}) leaves us with three independent parameters which are our initial conditions for the numerical shooting method. We choose the following parameters:
\bea \label{para}
\lambda^{(0)} \ \ , \ \ \chi^{(0)} \ \ , \ \ \Psi^{(1)} \ \ .
\eea
In the UV, i.e. near the AdS boundary $\tilde{z}=0$, the fields have the following expansion:
\begin{eqnarray}
&&\hskip -5.6cm \lambda(\tilde{z})=\lambda_{1}\tilde{z}+\lambda_{2}\tilde{z}^{2}+... \ , \nonumber \\
&&\hskip -5.6cm \chi(\tilde{z})=\chi_{0}+\lambda_{0}^{2}\tilde{z}^{2}+... \ , \nonumber \\
f(\tilde{z})=1+\lambda_{0}^{2}\tilde{z}^{2}+f_{3}\tilde{z}^{3}+... \ , \nonumber \\
&&\hskip -5.6cm \Psi(\tilde{z})=\Psi_{0}+\Psi_{1}\tilde{z}+... \label{uv} \ .
\end{eqnarray}
Using our initial conditions we first fix $\lambda^{(0)}$ and $\chi^{(0)}$. We then tune $\Psi^{(1)}$ so that either $\lambda_{1}=0$ or $\lambda_{2}=0$. In general this will generate some non-zero value for $\chi_{0}$ which we shift using the scaling symmetry to $\chi_{0}=0$. The UV asymptotic values of the field $\lambda$ define the vacuum expectation value of the charged operators in the theory and they are defined as:
\bea \label{lam}
\mathcal{O}_{1}=\frac{2\lambda_{1}}{\sqrt{16 \pi G_{4}}} \ \ , \ \ \mathcal{O}_{2}=\frac{2\lambda_{2}}{\sqrt{16 \pi G_{4}}R} \ .
\eea
Using the holographic dictionary, the UV asymptotics of the gauge field $\Psi(z)$ define a chemical potential $\mu$ and the charge density $\rho$ given by:
\bea \label{murho}
\mu =\frac{e^{\chi_{0}}/2}{\sqrt{16 \pi G_{4}}}\Psi_{0} \ \ , \ \ \rho =-\frac{e^{\chi_{0}}/2}{R\sqrt{16 \pi G_{4}}}\Psi_{1} \ .
\eea
The temperature can be computed in the usual way by Wick-rotating the metric (\ref{ans}) to Euclidean signature and then imposing regularity at the horizon \cite{Bobev:2011rv}:
\bea \label{temp}
T=\frac{1}{4\pi R \tilde{z}_{H}}\frac{e^{-(\chi^{(0)}-\chi_{0})/2}}{32}\Bigg(61+36\cosh \Big(2\lambda^{(0)}\Big)-\cosh \Big(4\lambda^{(0)}\Big)-8\tilde{z}^{2}_{H}e^{\chi^{(0)}}\Big(\Psi^{(1)}\Big)^{2} \Bigg) \ .
\eea
At high temperature the solution is the familiar RN--AdS black hole. At low enough temperature, below some critical value, there exists a new type of solution which has scalar hair. By computing the free energy in both cases it has been shown that the hairy black hole solution is thermodynamically preferred to that of the RN--AdS solution and the transition temperature $T_{c}$ has been obtained as well. See \cite{Bobev:2011rv},\cite{Albash:2012pd} for further details.

\subsection{RN--AdS Solution}
\label{rn}

The high temperature RN--AdS solution is obtained by setting $\lambda(z)=0$ and $\chi(z)=0$. That means both the operators $\mathcal{O}_{1}$ and $\mathcal{O}_{2}$ vanish and there is no condensate. The metric and the gauge field are given by:
\bea \label{rnsoln}
\Psi(z)=\frac{2QR}{z_{H}}\Bigg(1-\frac{z}{z_{H}}\Bigg) \  \ , \ \ f(z)=1-(1+Q^{2})\frac{z^{3}}{z^{3}_{H}}+Q^{2}\frac{z^{4}}{z^{4}_{H}} \ \ .
\eea
Using equations (\ref{murho}-\ref{temp}) the temperature, the chemical potential and the charge density become:
\bea \label{rntemp}
T=\frac{1}{4 \pi z_{H}}(3-Q^{2}) \ \ , \ \ \mu =\frac{R}{\sqrt{16 \pi G_{4}}}\frac{2Q}{z_{H}} \ \ , \ \ \rho =\frac{R}{\sqrt{16 \pi G_{4}}}\frac{2Q}{z_{H}^{2}} \ \ .
\eea

\subsection{Hairy Black Hole Solution}
\label{hairy}

As mentioned before, at low enough temperatures there exists a new branch of solutions which is a black hole with a charged scalar hair. There is no analytic solution available. So we employ a numerical shooting technique to obtain the solution. We impose the initial conditions in the IR and set $\chi^{(0)}=1$. Then by tuning $\Psi^{(1)}$ we set either $\lambda_{1}=0$ or $\lambda_{2}=0$, meaning, either $\mathcal{O}_{2}$ or $\mathcal{O}_{1}$ being turned on respectively. Below the critical temperature $T_{c}$ this new type of solution represents the superconducting phase with non--zero condensate.

\subsection{Zero Temperature Solution}
\label{zero}

It is argued in \cite{Gubser:2008wz} that the zero temperature solution is an RG flow between two AdS$_{4}$ spaces. We will again use the numerical shooting technique and impose the initial conditions in the IR. Since there is no black hole horizon at zero temperature, the IR is now at $\tilde{z}\rightarrow \infty$. In the IR, the fields have an expansion of the form \cite{Bobev:2011rv}:
\begin{eqnarray}
&&\hskip -0.5cm \lambda (\tilde{z})=\log (2+\sqrt{5})+\lambda^{1}\tilde{z}^{-\alpha}+... \ , \nonumber \\
&&\hskip -0.5cm \Psi(\tilde{z})=\psi^{1}\tilde{z}^{-\beta}+... \ , \nonumber \\
&&\hskip -0.5cm f(\tilde{z})=\frac{7}{3}+... \ , \nonumber \\ 
&&\hskip -0.5cm \chi(\tilde{z})=\chi^{0}+... \ \label{zerosol},
\end{eqnarray}
where,
\bea
\alpha = \sqrt{\frac{303}{28}}-\frac{3}{2} \ , \ \ \beta =\sqrt{\frac{247}{28}}-\frac{1}{2} \ .
\eea
As before, using the scaling symmetries (\ref{sym}) of the equations of motion, we can fix the values of $\Psi^{1}$ and $\chi^{0}$ and then tune the free parameter $\lambda^{1}$ to either have $\lambda_{1}=0$ or $\lambda_{2}=0$ in the UV. We set $\Psi^{1}=1$ and $\chi^{0}=4$.

\section{The Entanglement Entropy and The Subregion Complexity}
\label{eehc}

It is instructive to reproduce the results of the entanglement entropy reported in \cite{Albash:2012pd} for completeness and also so that we can compare that with our subregion complexity results in the next section. To that end, let us choose a strip region $\mathcal{A}$ with width $l$ and length $L\rightarrow \infty$ in a constant
\begin{figure}[h!]
   \begin{center}
{\centering
\includegraphics[width=3.5in]{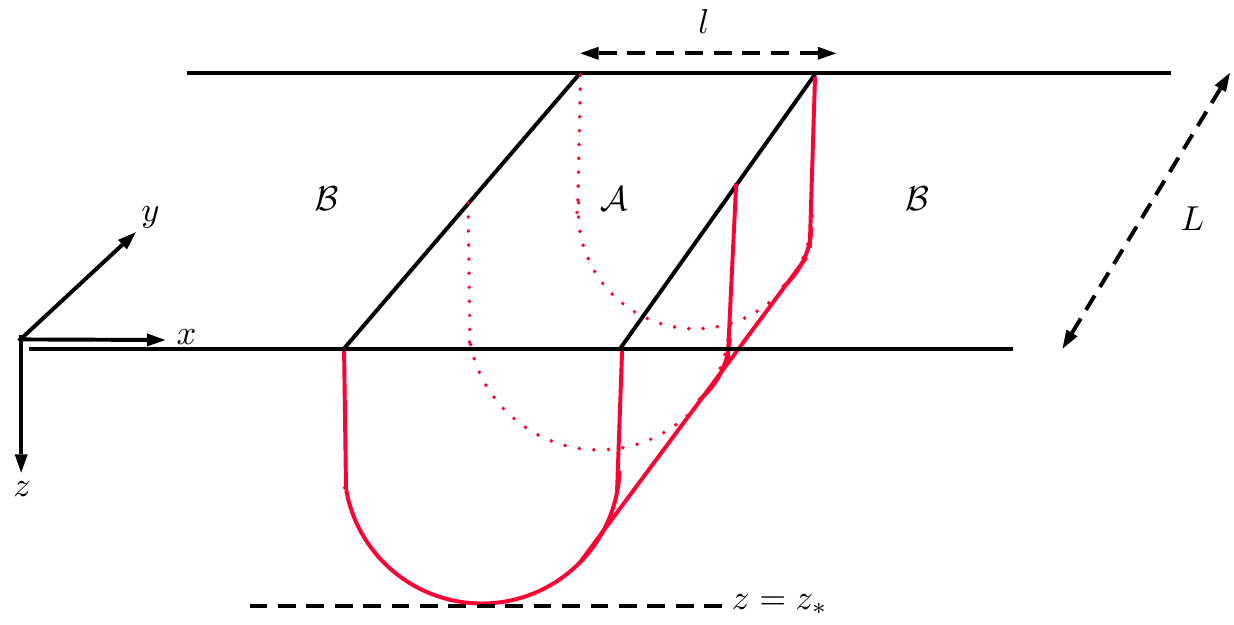} 
   \caption{\footnotesize  The strip geometry considered in this paper. Here $z$ denotes the radial direction in the dual bulk geometry AdS$_{4}$. The strip width is $l$ and the length is $L$ which can be taken to infinity. The boundary is at $z=0$ where the field theory lives and $z=z_{*}$ denotes the turning point of the minimal surface inside the bulk.}   \label{fig:strip}
}
   \end{center}
\end{figure}
time slice (figure~\ref{fig:strip}). Now following the RT proposal, we need to find the minimal surface $\gamma_{\mathcal{A}}$ bounded by the perimeter of $\mathcal{A}$ and that extends into the bulk of the geometry. Then the area of this minimal surface will give us the entanglement entropy of the subregion $\mathcal{A}$ using equation (\ref{eeform}) and the volume enclosed by this minimal surface and the strip region $\mathcal{A}$ will give us the subregion complexity using equation (\ref{cvform}). Since our background is static we can compute this volume by slicing the bulk with planes of constant $z$. See \cite{Ben-Ami:2016qex} for a review on the computation of volumes of subregions in various AdS black hole geometries. In ref. \cite{Carmi:2016wjl} the authors have further discussed the divergence structure of the volume and a covariant generalization of computing the volume which includes time-dependent geometries as well.
The $2$-dimensional minimal surface is given by minimizing the area functional: 
\bea \label{area}
{\rm Area}(\gamma_{\mathcal{A}})=L\int^{l/2}_{-l/2} dx\frac{R^{2}}{z^{2}}\Bigg(1+\frac{z^{\prime}(x)^{2}}{f(z)}\Bigg)^{1/2} \ .
\eea
The minimization problem leads to the Hamiltonian which is independent of $x$ and hence this is a constant of motion:
\bea \label{com}
\frac{1}{z^{2}_{*}}=\frac{1}{z^{2}}\frac{1}{\sqrt{1+\frac{z^{\prime}(x)^{2}}{f(z)}}} \ ,
\eea
where $z=z_{*}$ is the turning point of the minimal surface in the bulk. This equation determines the minimal surface $\gamma_{A}$:
\bea \label{mini}
\frac{dz}{dx}=\frac{\sqrt{(z^{4}_{*}-z^{4})f(z)}}{z^{2}} \ .
\eea
Since $z=z_{*}$ is the turning point of the minimal surface in the bulk we will require that:
\bea \label{width}
\frac{l(z_{*})}{2}=\int^{z_{*}}_{0}dz\frac{z^{2}}{\sqrt{(z^{4}_{*}-z^{4})f(z)}} \ .
\eea
Then using (\ref{eeform}) and (\ref{com}) or (\ref{mini}), the entanglement entropy becomes:
\bea \label{ent}
4G_{4}S=2LR^{2}\int^{z_{*}}_{\epsilon}dz\frac{z^{2}_{*}}{z^{2}}\frac{1}{\sqrt{(z^{4}_{*}-z^{4})f(z)}}=2LR^{2}\Bigg(s+\frac{1}{\epsilon}\Bigg) \ ,
\eea
where $s$ is the finite part and has dimension of inverse length \cite{Albash:2012pd}. Note that we have introduced a small cut-off $\epsilon$ to regularize the area integral.\\
To compute the subregion complexity, we need to find out the volume enclosed by the minimal surface $\gamma_{A}$ and the strip region $\mathcal{A}$. This can easily be done by integrating the inside of the minimal surface:
\bea \label{vol}
V(z_{*})=2LR^{3}\int^{z_{*}}_{\epsilon}dz\frac{1}{z^{3}\sqrt{f(z)}}x(z) \ ,
\eea
where, 
\bea \label{one}
x(z)=\int^{z_{*}}_{z}du\frac{u^{2}}{\sqrt{(z^{4}_{*}-u^{4})f(u)}} \ .
\eea
Then using (\ref{cvform}) and (\ref{vol}) the subregion complexity becomes:
\bea \label{comp}
8\pi GR\mathcal{C}=2LR^{3}\Bigg(c_{fin}+\frac{x(0)}{2\epsilon^{2}}\Bigg) \ ,
\eea
where, $c_{fin}$ is the finite part of the subregion complexity and has dimension of inverse length. Note that, 
\bea
x(0)=\int^{z_{*}}_{0}dz\frac{z^{2}}{\sqrt{(z^{4}_{*}-z^{4})f(z)}}\equiv \frac{l(z_{*})}{2} \ .
\eea
Since the diverging term has $z_{*}$ dependence we should divide the quantity in the big parenthesis in equation (\ref{comp}) by $x(0)$ and then plot this re-scaled finite complexity $c$ as a function of the strip width $l$ or the temperature $T$. This ensures that we subtract the same diverging term for each $z_{*}$. Finally, symmetries allow us to use the following dimensionless quantities to analyze our system:
\bea
\frac{T}{\sqrt{\rho}} \ , \ \frac{\mathcal{O}_{1}}{\sqrt{\rho}} \ , \ \frac{\mathcal{O}_{2}}{\rho} \ , \ \sqrt{\rho} \ l \ , \ \frac{s}{\sqrt{\rho}} \ , \ \frac{c}{\rho} \ .
\eea

\subsection{$\mathcal{O}_{1}$ Superconductor}
\label{o1}

\begin{figure}[h]
\centering
\begin{subfigure}[b]{0.3\textwidth}
\includegraphics[width=2.5in]{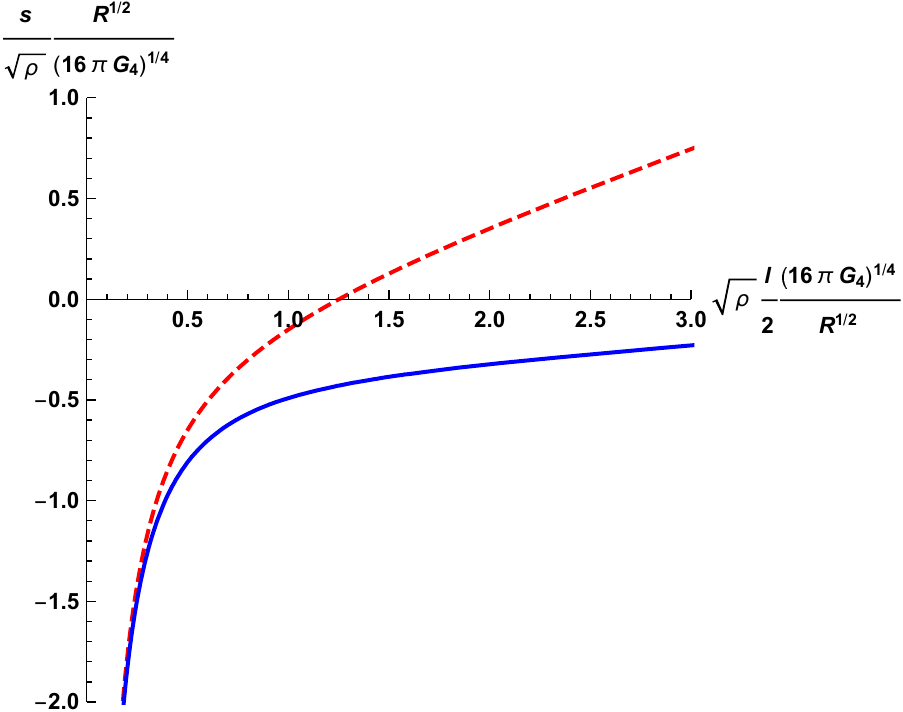} 
   \caption{\footnotesize  }   
 \label{fig:o1sl}
\end{subfigure}
\qquad\qquad
\begin{subfigure}[b]{0.3\textwidth}
\includegraphics[width=2.5in]{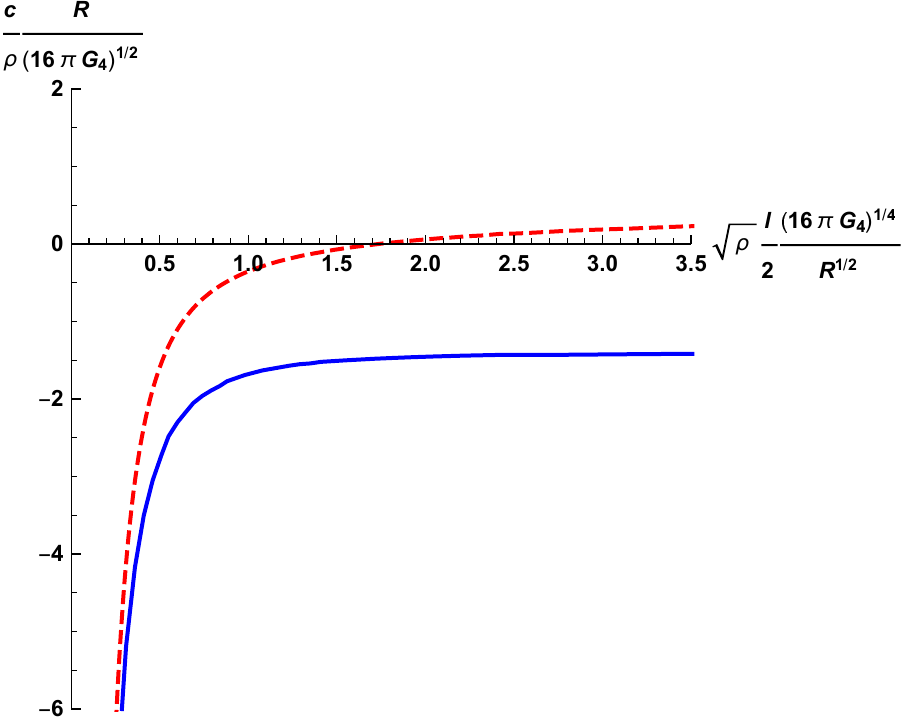} 
   \caption{\footnotesize  }   
\label{fig:o1cl}
\end{subfigure}
   \caption{\footnotesize The entanglement entropy (a) and the subregion complexity (b) as functions of the strip--width $l$ for the $\mathcal{O}_1$ superconductor for a fixed temperature: $\frac{R^{1/2}}{(16\pi G_{4})^{1/4}}\frac{T}{\sqrt{\rho}}=0.053$. The red dashed curve is the Reissner--Nordstrom solution and the solid blue curve is the superconductor solution.}   
   \label{fig:o1scl}
\end{figure}
Figure~\ref{fig:o1scl} shows our results for the entanglement entropy $s$ and the subregion complexity $c$ as functions of the strip width $l$ for a fixed temperature: $\frac{R^{1/2}}{(16\pi G_{4})^{1/4}}\frac{T}{\sqrt{\rho}}=0.053$, which is below the transition temperature $T_{c}$. In both cases we see the expected linear growth behavior for large $l$ (for the entanglement entropy this is known as the ``area law'') and we find that the entropy and the complexity are lower in the superconducting phase than that of the normal phase.
\begin{figure}[h!]
\centering
\begin{subfigure}[b]{0.3\textwidth}
\includegraphics[width=2.5in]{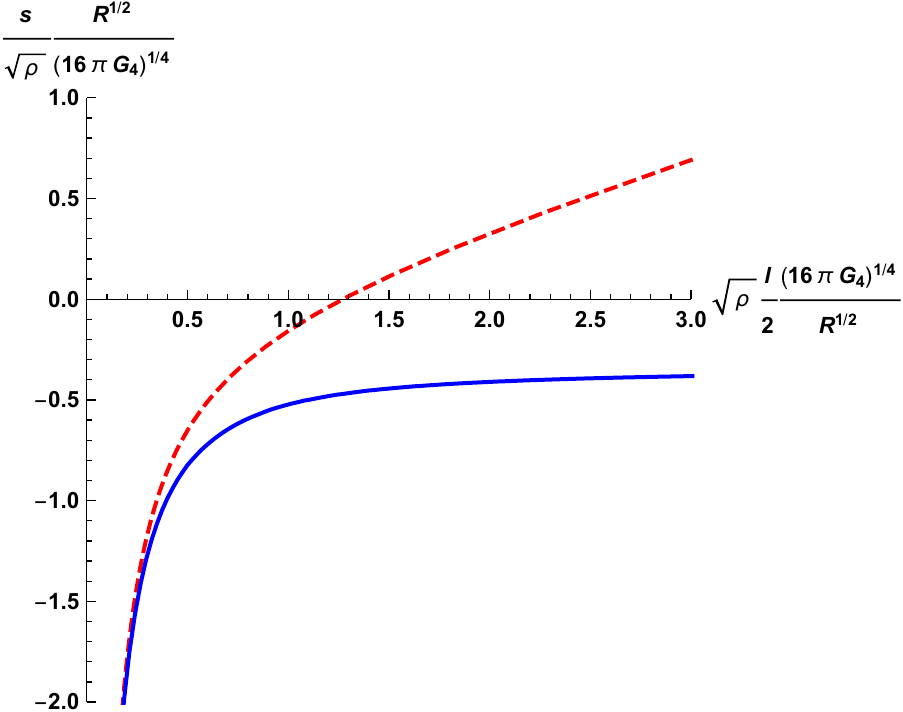} 
   \caption{\footnotesize  }   
 \label{fig:zo1sl}
\end{subfigure}
\qquad\qquad
\begin{subfigure}[b]{0.3\textwidth}
\includegraphics[width=2.5in]{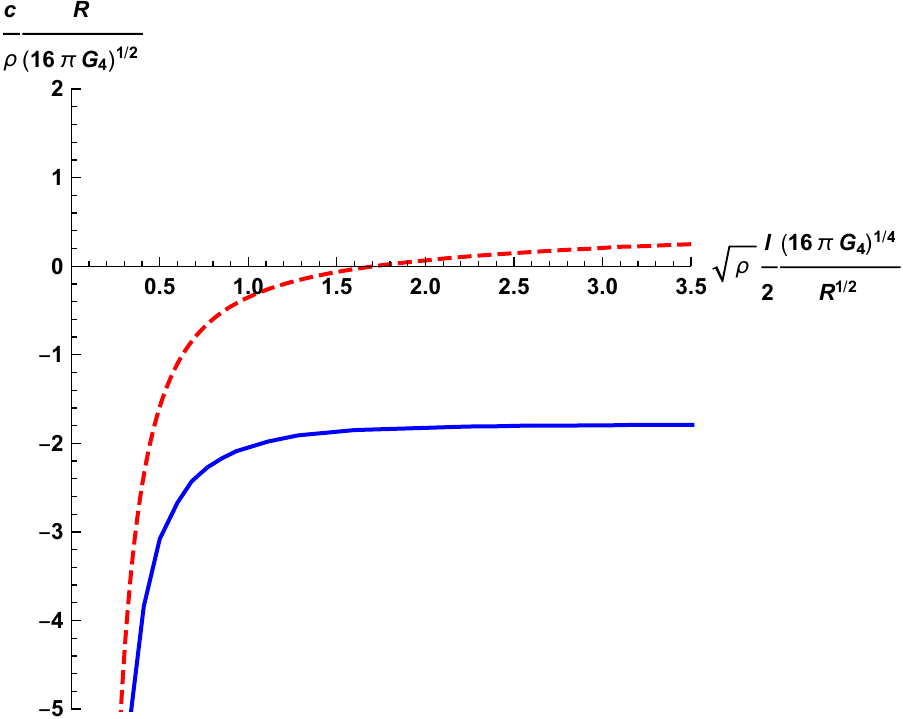} 
   \caption{\footnotesize  }   
\label{fig:zo1cl}
\end{subfigure}
   \caption{\footnotesize The entanglement entropy (a) and the subregion complexity (b) as functions of the strip--width $l$ for the $\mathcal{O}_1$ superconductor for a fixed temperature: $\frac{R^{1/2}}{(16\pi G_{4})^{1/4}}\frac{T}{\sqrt{\rho}}=0$. The red dashed curve is the Reissner--Nordstrom solution and the solid blue curve is the superconductor solution.}   
   \label{fig:zo1scl}
\end{figure}
\begin{figure}[h!]
\centering
\begin{subfigure}[b]{0.3\textwidth}
\includegraphics[width=2.5in]{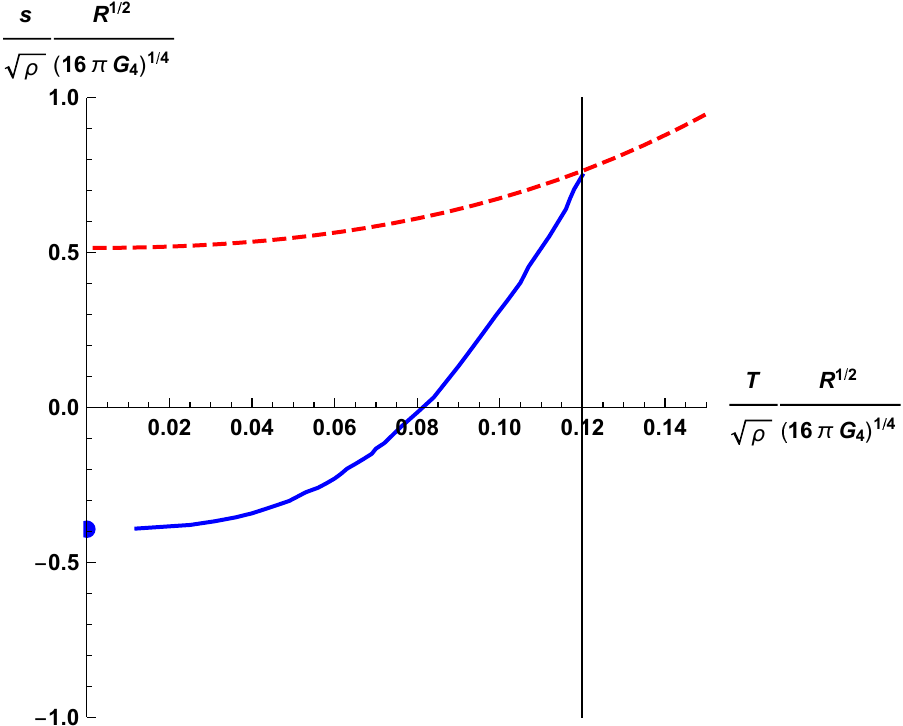} 
   \caption{\footnotesize  }   
 \label{fig:o1st}
\end{subfigure}
\qquad\qquad
\begin{subfigure}[b]{0.3\textwidth}
\includegraphics[width=2.5in]{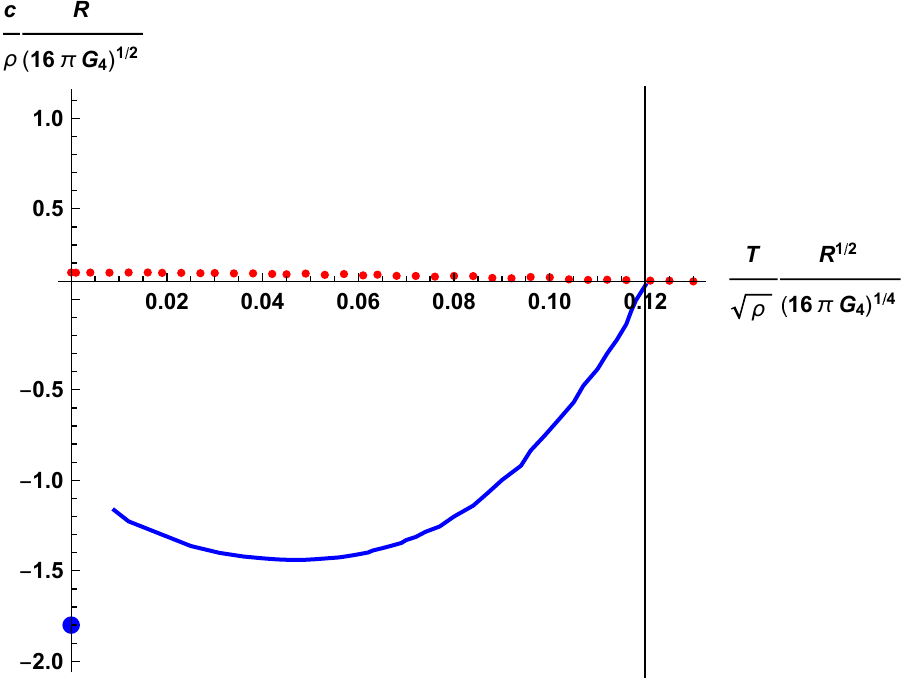} 
   \caption{\footnotesize  }   
\label{fig:o1ct}
\end{subfigure}
   \caption{\footnotesize The entanglement entropy (a) and the subregion complexity (b) as functions of the temperature $T$ for the $\mathcal{O}_1$ superconductor for a fixed $l$: $\sqrt{\rho}(16\pi G_{4})^{1/4}R^{-1/2}l/2=2.5$. The red dashed(or dotted) curve is the Reissner--Nordstrom solution and the solid blue curve is the superconductor solution. The black solid line denotes the transition temperature $T_{c}$. Since the zero temperature solution is exactly known, we include our results for $T=0$ in the plot.}   
   \label{fig:o1sct}
\end{figure}
The RN--AdS case having higher entropy than the superconducting case represents the fact that the degrees of freedom have condensed in the latter case. As we decrease the temperature all the curves still go linearly for large $l$ though the slopes of the curves in superconducting cases are smaller. In figure~\ref{fig:zo1scl} we plot our results for $T=0$. Remember that the zero temperature solution is an RG flow between two AdS vacua \cite{Gubser:2008wz, Bobev:2011rv}. Now, large $l$ probes more deeply into the IR and for empty AdS since there is no horizon, the IR is at $\tilde{z}\rightarrow \infty$ where $f(\tilde{z})$ is constant. Hence, the entropy and the complexity approaching different constant values for large $l$ during the superconducting phase is not surprising.

In figure~\ref{fig:o1sct} we present how $s$ and $c$ change with the temperature while the strip width $l$ is kept fixed. For the entropy plot, the physical curve is determined by choosing the point of lowest entropy for a given temperature \cite{Albash:2012pd}. As we lower the temperature the entropy decreases in both the phases and there is a discontinuity in the slope at the transition temperature $T_{c}$. On the contrary, as we lower the temperature the complexity increases during the normal phase but decreases during the superconducting phase. At some low temperature, the complexity in the superconducting phase rises slightly and then drops to a finite minimum value at zero temperature. Note that we do not plot all the superconductor results due to lack of numerical control in our shooting technique. Similar to the entropy plot, there is again a discontinuity in the slope at the transition temperature $T_{c}\approx 0.1199\frac{\sqrt{\rho}(16\pi G_{4})^{1/4}}{R^{1/2}}$. Also note that both plots lead to the same transition temperature $T_{c}$.

\subsection{$\mathcal{O}_{2}$ Superconductor}
\label{o2}

\begin{figure}[h]
\centering
\begin{subfigure}[b]{0.3\textwidth}
\includegraphics[width=2.5in]{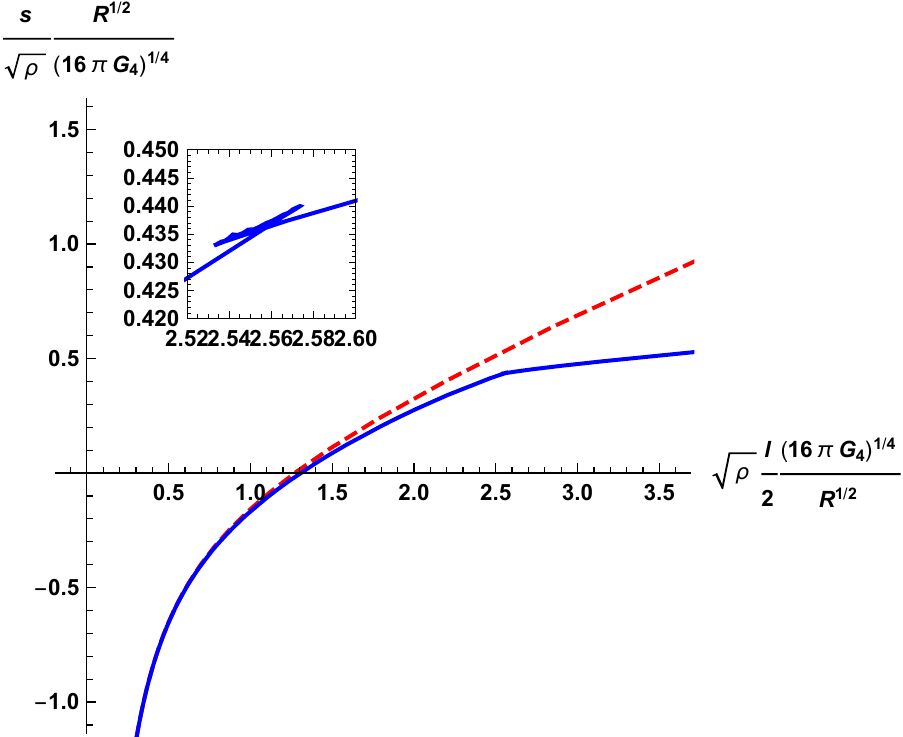} 
   \caption{\footnotesize  }   
 \label{fig:o2sl}
\end{subfigure}
\qquad\qquad
\begin{subfigure}[b]{0.3\textwidth}
\includegraphics[width=2.5in]{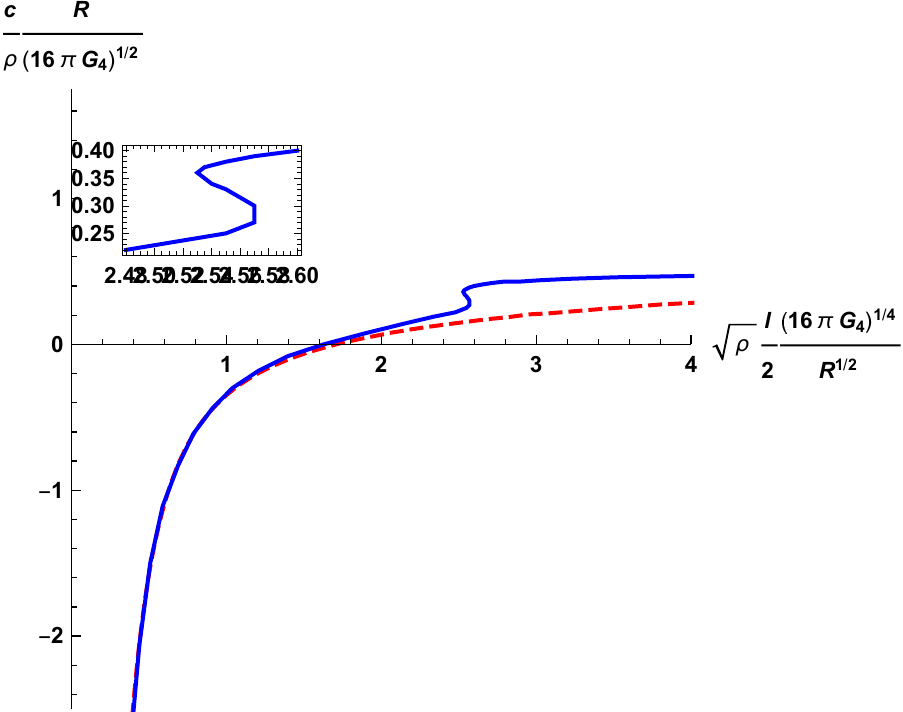} 
   \caption{\footnotesize  }   
\label{fig:o2cl}
\end{subfigure}
   \caption{\footnotesize The entanglement entropy (a) and the subregion complexity (b) as functions of the strip--width $l$ for the $\mathcal{O}_2$ superconductor for a fixed temperature: $\frac{R^{1/2}}{(16\pi G_{4})^{1/4}}\frac{100T}{\sqrt{\rho}}=0.305$. The red dashed curve is the Reissner--Nordstrom solution and the solid blue curve is the superconductor solution.}   
   \label{fig:o2scl}
\end{figure}
\begin{figure}[h]
\centering
\begin{subfigure}[b]{0.3\textwidth}
\includegraphics[width=2.5in]{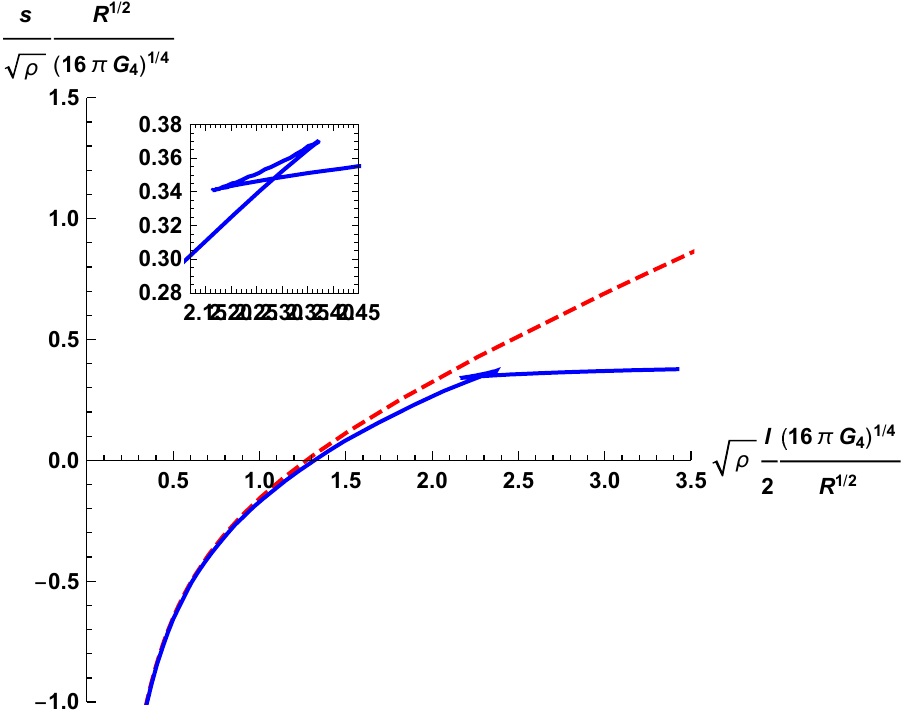} 
   \caption{\footnotesize  }   
 \label{fig:zo2sl}
\end{subfigure}
\qquad\qquad
\begin{subfigure}[b]{0.3\textwidth}
\includegraphics[width=2.5in]{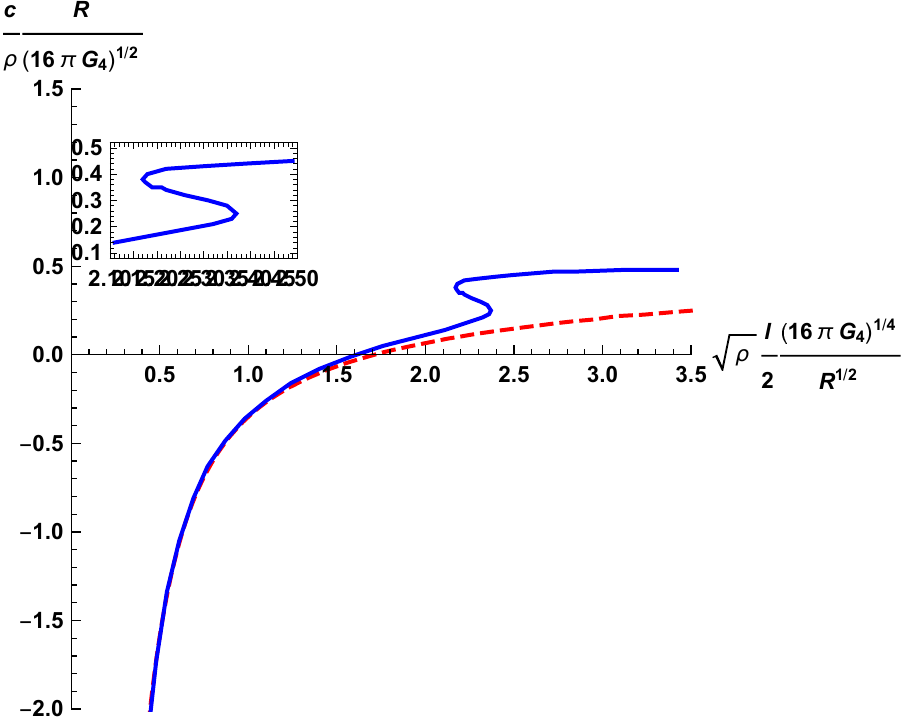} 
   \caption{\footnotesize  }   
\label{fig:zo2cl}
\end{subfigure}
   \caption{\footnotesize The entanglement entropy (a) and the subregion complexity (b) as functions of the strip--width $l$ for the $\mathcal{O}_2$ superconductor for a fixed temperature: $\frac{R^{1/2}}{(16\pi G_{4})^{1/4}}\frac{100T}{\sqrt{\rho}}=0$. The red dashed curve is the Reissner--Nordstrom solution and the solid blue curve is the superconductor solution.}   
   \label{fig:zo2scl}
\end{figure}
Figure~\ref{fig:o2scl} again shows our results for the entanglement entropy $s$ and the subregion complexity $c$ as functions of the strip width $l$ for a fixed temperature: $\frac{R^{1/2}}{(16\pi G_{4})^{1/4}}\frac{100T}{\sqrt{\rho}}=0.305$. We choose the fixed temperature to be below the transition temperature $T_{c}$ as before. For $\mathcal{O}_{2}$ superconductor the critical temperature is given by: $T_{c}\approx 0.3638\frac{\sqrt{\rho}(16\pi G_{4})^{1/4}}{R^{1/2}}$. In \cite{Albash:2012pd}, the origin of this multivaluedness has been argued to be the non--monotonic behavior of $f(z)$, which applies to the results of the complexity as well. In figure~\ref{fig:zo2scl} we plot our results for $T=0$. In general both the plots behave similar to that of the $\mathcal{O}_{1}$ case for large $l$. But there are few crucial differences here now. For the temperature below the critical temperature $T_{c}$ both the quantities now show multivaluedness for a given range of $l$. For the entropy this fact is reflected in the form of a swallowtail \cite{Albash:2010mv, Albash:2012pd}. In the complexity plot this is captured as an ``S'' curve. Also note that the superconducting phase now has higher complexity than the normal phase in the region of multivaluedness and for large $l$ as well which is in contrast with the entropy plot.
\begin{figure}[h]
\centering
\begin{subfigure}[b]{0.3\textwidth}
\includegraphics[width=2.2in]{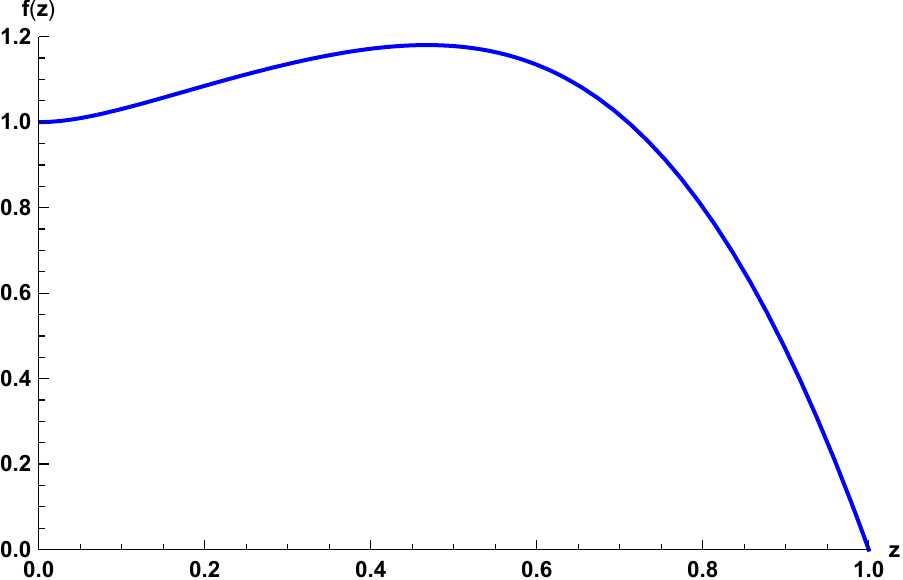} 
   \caption{\footnotesize  }   
 \label{fig:o1f}
\end{subfigure}
\qquad\qquad
\begin{subfigure}[b]{0.3\textwidth}
\includegraphics[width=2.2in]{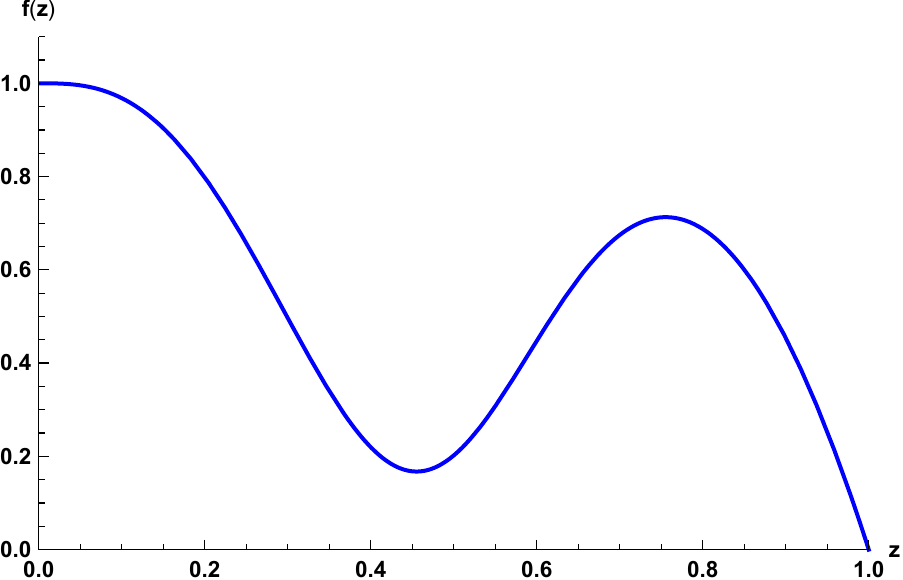} 
   \caption{\footnotesize  }   
\label{fig:o2f}
\end{subfigure}
   \caption{\footnotesize (a) Behavior of $f(z)$ for the $\mathcal{O}_{1}$ superconductor. This has been shown for: $\frac{R^{1/2}}{(16\pi G_{4})^{1/4}}\frac{T}{\sqrt{\rho}}=0.053$. (b) Behavior of $f(z)$ for the $\mathcal{O}_{2}$ superconductor. This has been shown for: $\frac{R^{1/2}}{(16\pi G_{4})^{1/4}}\frac{100T}{\sqrt{\rho}}=0.305$.}   
   \label{fig:eff}
\end{figure}
It has been observed that in the $\mathcal{O}_{2}$ case, $f(z)$ can develop a minimum and a maximum (at low temperatures). When the turning point of the minimal RT surface in the bulk, $z_{*}$ lies in the neighborhood of the minimum of $f(z)$, the entropy and the complexity become multivalued. See \cite{Albash:2012pd} for further details and a clear demonstration of how it happens using the domain wall analysis.
\begin{figure}[h!]
\centering
\begin{subfigure}[b]{0.25\textwidth}
\includegraphics[width=2.0in]{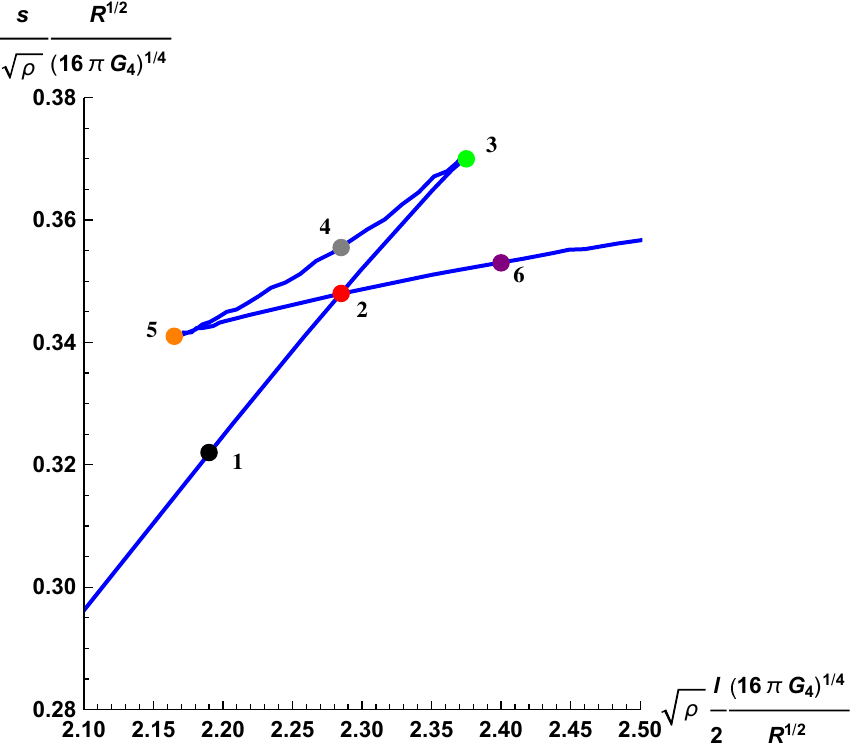} 
   \caption{\footnotesize  }   
 \label{fig:cs}
\end{subfigure}
\qquad\qquad
\begin{subfigure}[b]{0.25\textwidth}
\includegraphics[width=2.0in]{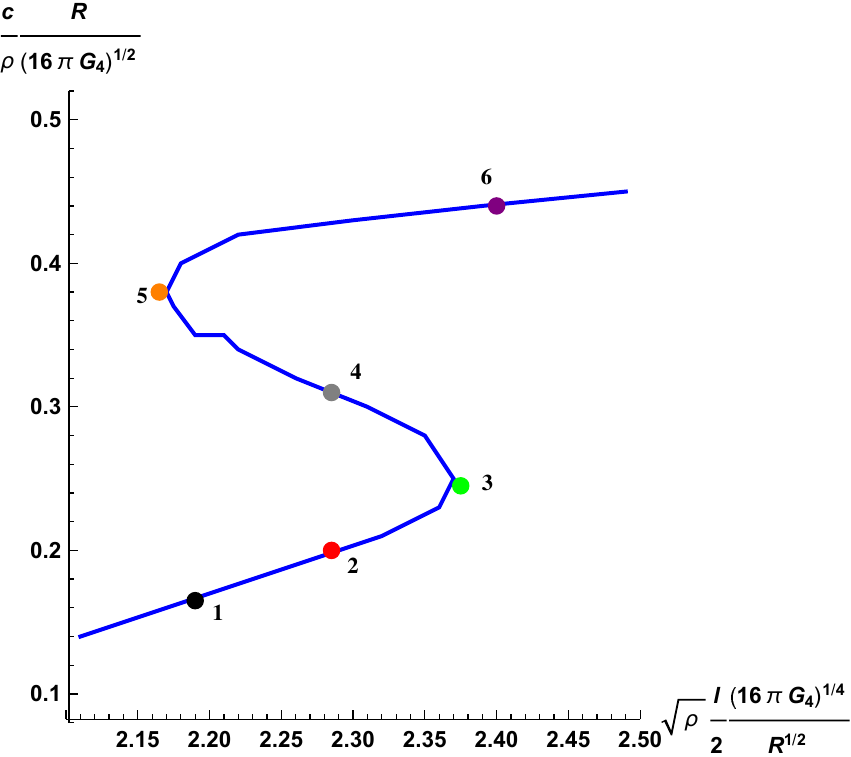} 
   \caption{\footnotesize  }   
\label{fig:cc}
\end{subfigure}
   \caption{\footnotesize A comparison of the multivalued regions of the entanglement entropy (a) and the subregion complexity (b) as functions of the strip--width $l$ for the $\mathcal{O}_2$ superconductor for a fixed temperature: $\frac{R^{1/2}}{(16\pi G_{4})^{1/4}}\frac{T}{\sqrt{\rho}}=0$.}   
   \label{fig:compsc}
\end{figure}
We show the behavior of $f(z)$ in figure~\ref{fig:eff}. In figure~\ref{fig:compsc} we compare different parts of the entanglement entropy and the subregion complexity curves for zero temperature by mapping out the corrsponding features of the multivalued regions. From the entropy plot (figure~\ref{fig:cs}) we see that the physical curve must follow the sequence $6\rightarrow 2\rightarrow 1$ as explained in \cite{Albash:2012pd}. Accordingly, there is a finite jump towards a lower value in the complexity plot (figure \ref{fig:cc}) from $6\rightarrow 2\rightarrow 1$.
\begin{figure}[h!]
\centering
\begin{subfigure}[b]{0.3\textwidth}
\includegraphics[width=2.5in]{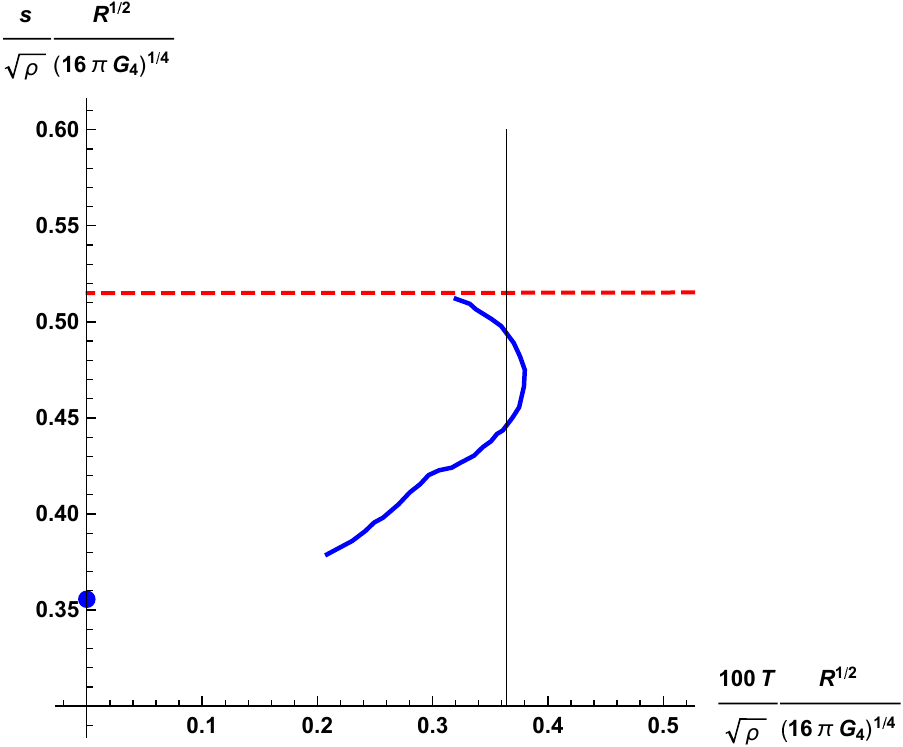} 
   \caption{\footnotesize  }   
 \label{fig:o2st}
\end{subfigure}
\qquad\qquad
\begin{subfigure}[b]{0.3\textwidth}
\includegraphics[width=2.5in]{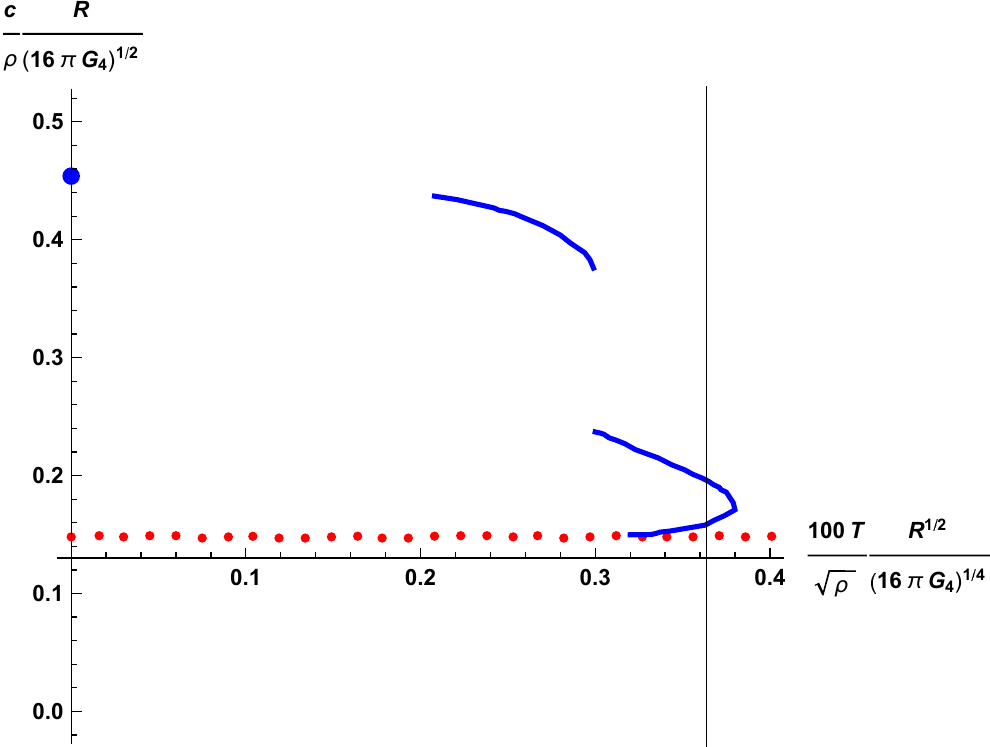} 
   \caption{\footnotesize  }   
\label{fig:o2ct}
\end{subfigure}
   \caption{\footnotesize The entanglement entropy (a) and the subregion complexity (b) as functions of the temperature $T$ for the $\mathcal{O}_2$ superconductor for a fixed $l$: $\sqrt{\rho}(16\pi G_{4})^{1/4}R^{-1/2}l/2=2.5$. The red dashed(or dotted) curve is the Reissner--Nordstrom solution and the solid blue curve is the superconductor solution. The black solid line denotes the transition temperature $T_{c}$. Since the zero temperature solution is exactly known, we include our results for $T=0$ in the plot.}   
   \label{fig:o2sct}
\end{figure}

We present in figure~\ref{fig:o2sct} how $s$ and $c$ change with the temperature while the strip width $l$ is kept fixed. There is again a discontinuity in the slope at the transition temperature $T_{c}$ in both the cases. Moreover, as we lower the temperature, the value of the entropy drops discontinuously whereas the value of the complexity rises discontinuously. There is another special feature in this case. For our chosen value of $l$,  there is an additional discontinuity in the slope of the decreasing entropy at some lower temperature and the entropy curve is a combination of two types of curve joined by this new discontinuity. This special feature is due to a new length scale in the theory as argued in \cite{Albash:2012pd}. In the complexity plot, correspondingly we see a discontinuous but finite jump  exactly at the temperature where the above-mentioned new discontinuity shows up in the entropy plot.

\section{Summary and Outlook}
\label{sum}

In summary, we have performed a numerical shooting technique to compute the subregion complexity for a $2+1$--dimensional holographic superconductor using the ``Complexity equals Volume'' or the CV conjecture. Our analysis reveals that apart from the universal divergent term, there are no other divergences in the complexity as far as the phase transition is concerned. The subregion complexity grows linearly with $l$ for large strip width $l$. From our computation it is clear that the complexity captures phase transition as well and the transition temperature $T_{c}$ is exactly same as computed from the free energy analysis of the system. Same $T_{c}$ has been read off from the entropy vs. temperature plot as well. This is not surprising since the gravity background is dual to a field theory with a single transition temperature and hence our result is a nice confirmation of holography.\footnote{We thank the anonymous referee for pointing this out.}Apart from that, the complexity actually behaves differently. Specially, for the first order phase transition, the complexity of the superconducting phase is higher than the normal phase, and it also increases with decreasing temperature. The second order phase transition rather shows a similar behavior to that of the entropy analysis, though, at very low temperature the complexity behavior is quite strange and the physics behind it is yet to be investigated. We have shown the zero temperature solution in all these cases as well. We have also observed the multivaluedness similar to the entropy plot except the fact that in this case the multivaluedness is of different form, namely in the form of an ``S'' curve. Finally, the behavior of the complexity suggests that it may be used as another independent probe to the physics of the phase transition.

Our results are in agreement with that reported in \cite{Zangeneh:2017tub} and in \cite{Fujita:2018xkl} where the authors have studied $1+1$--dimensionsal $s$-wave and $p$-wave holographic superconductor respectively: the complexity remains finite during the phase transition\footnote{Notice that our results do not match with the results reported in \cite{Momeni:2016ekm}. They have found that during the phase transition the complexity becomes infinite for a $1+1$-dimensional $s$-wave superconductor.} and the subregion complexity plot leads to the same transition temperature as observed from the entropy plot. Ref. \cite{Fujita:2018xkl} has also observed multivaluedness and discontinuous but finite jump in the complexity during the first order phase transition. While this project was near completion two new articles appeared in the literature: In \cite{Yang:2019gce} the authors have studied the time dependent complexity and how the complexity (of formation) scales with the temperature where the scaling factor is a function of the superconductor model parameters by using the CV conjecture in an asymptotically AdS$_{d+1}$ geometry. In \cite{Guo:2019vni} the authors have investigated the subregion complexity for the St\"{u}ckelberg superconductor which is very similar to our set up in this paper. Apart from the second order phase transition result, all other main results reported here agree with theirs as well. During the second order phase transition the complexity of the superconducting phase is opposite to what they have found.\footnote{See also \cite{Roy:2017kha} for analytic expressions of the complexity in the high and the low temperature regime for the Schwarzschild--AdS and the RN--AdS black holes.}\footnote{In refs. \cite{Bhattacharya:2018oeq} the authors have calculated holographic entanglement entropy, subregion complexity and fisher information metric for a class of nonsupersymmetric D3 branes and also reviewed the same for the AdS black brane as well.}  

There are multiple questions which need to be answered. It has been reported that the complexity decreases with increasing temperature during the second order phase transition, but in this paper we actually found that it behaves quite similar to that of the entropy for the superconducting phase. It seems like the temperature dependence of the complexity is not universal. It will be nice to see if there is any deeper physics behind it. Also, the behavior of the complexity of the $\mathcal{O}_1$ superconductor at low temperature needs some careful explanation. Some other questions that we might ask is what happens if we compute the complexity using the CA conjecture and how the complexity evolves after a thermal quench? Finally, what all of these mean in the the dual field theory is worth exploring. We will try to answer some of these in future work.

\section*{Acknowledgments}

We would like to thank the  US Department of Energy for partial support under grant de-sc 0011687. AC thanks Clifford V. Johnson, Tameem Albash, Nicholas P. Warner, Siavash Yasini, Arnab Kundu, Chethan Krishnan, P. N. Bala Subramanian and Felipe Rosso for helpful discussions and comments.


\appendix
\section{The Subregion Complexity in (d+1)--dimensions}
\label{gen}

Let us consider a general asymptotically AdS$_{d+1}$ metric:
\bea
ds^{2}=\frac{1}{z^{2}}[f_{0}(z)dt^{2}+f_{1}(z)dx^{2}_{\mu}+f_{2}(z)dz^{2}] \ ,
\eea
where $f_{0}(z)$, $f_{1}(z)$ and $f_{2}(z)$ are some arbitrary functions of $z$ with $f_{i}(z=0)=1$. We choose the entangling region as a strip of width $l$ and length $L\rightarrow \infty$. The entanglement entropy then turns out to be the following \cite{Ben-Ami:2016qex}:
\bea
S(z_{*})=\frac{2L^{d-2}}{4G_{N}}\int^{z_{*}}_{\epsilon}\frac{dz}{z^{d-1}}\frac{\sqrt{f_{2}f^{d-1}_{1}}}{\sqrt{1-\frac{f^{d-1}_{1}(z_{*})z^{2d-2}}{f^{d-1}_{1}(z)z^{2d-2}_{*}}}} \ ,
\eea
and the subregion complexity is given by:
\bea
\mathcal{C}(z_{*})=\frac{2L^{d-2}}{8\pi GR}\int^{z_{*}}_{\epsilon}\frac{dz\sqrt{f^{d-1}_{1}f_{2}}}{z^{d}}x(z) \ ,
\eea
where,
\bea
x(z)=\int^{z_{*}}_{z}dZ \frac{\sqrt{\frac{f_{2}(Z)}{f_{1}(Z)}}}{\sqrt{\frac{f^{d-1}_{1}(Z)z^{2d-2}_{*}}{f^{d-1}_{1}(z_{*})Z^{2d-2}}-1}} \ .
\eea
As usual, $z=z_{*}$ is the turning point of the minimal surface inside the bulk and $\epsilon$ is a small UV cut-off. The strip width $l$ is:
\bea
\frac{l(z_{*})}{2}=\int^{z_{*}}_{\epsilon}dz \frac{\sqrt{\frac{f_{2}(z)}{f_{1}(z)}}}{\sqrt{\frac{f^{d-1}_{1}(z)z^{2d-2}_{*}}{f^{d-1}_{1}(z_{*})z^{2d-2}}-1}} \ .
\eea
If we choose: $f_{0}=-R^{2}f(z)e^{-\chi(z)}$, $f_{1}=R^{2}$, $f_{2}=\frac{R^{2}}{f(z)}$ and $d=3$, we arrive at our results (\ref{width}-\ref{comp}).


%
\bibliographystyle{utphys}
\bibliography{AvikChak}




\end{document}